\documentclass[conference]{IEEEtran}
\usepackage{cite}
\usepackage[pdftex]{graphicx}
\usepackage[pdftex]{color}
\usepackage[ruled,vlined,linesnumbered]{algorithm2e}
\usepackage[cmex10]{amsmath}
\usepackage{algpseudocode}
\usepackage{array}
\usepackage{url}
\usepackage{bm} 
\usepackage{amssymb}
\usepackage{amsthm}
\usepackage{booktabs}

\theoremstyle{remark}

\theoremstyle{definition}

\begin{document}
\title{Massive M2M Access with Reliability Guarantees in LTE Systems}
\author{\IEEEauthorblockN{Germ\'an Corrales Madue\~no,  Nuno K. Pratas,   \v Cedomir Stefanovi\' c, Petar Popovski}
\IEEEauthorblockA{\\Department of Electronic Systems, Aalborg University, Denmark \\
Email: \{gco,nup,cs,petarp\}@es.aau.dk}}

\maketitle

\begin{abstract}
	Machine-to-Machine (M2M) communications are one of the major drivers of the cellular network evolution towards 5G systems. One of the key challenges is on how to provide reliability guarantees to each accessing device in a situation in which there is a massive number of almost-simultaneous arrivals from a large set of M2M devices. The existing solutions take a \emph{reactive} approach in dealing with massive arrivals, such as non-selective barring when a massive arrival event occurs, which implies that the devices cannot get individual reliability guarantees. In this paper we propose a \emph{proactive} approach, based on a standard operation of the cellular access. The access procedure is divided into two phases, an estimation phase and a serving phase. In the estimation phase the number of arrivals is estimated and this information is used to tune the amount of resources allocated in the serving phase. Our results show that the proactive approach is instrumental in delivering high access reliability to the M2M devices. 
\end{abstract}

\section{Introduction}
\label{sec:Introduction}
	Among the major drivers for the evolution of current cellular networks towards the fifth generation (5G) is the efficient support of Machine-to-Machine (M2M) communications and services.
Different from human-centric services (H2x), which are mainly characterized by the ever-increasing data rates, M2M services pose a different set of challenges, associated with the support of a massive number of users exchanging small amounts of data, often with requirements in terms of reliability and availability. 
A model for a particularly demanding M2M scenario is the one where the cellular network access should be offered with reliability guarantees in the case of massive almost-simultaneous arrivals.
An example is correlated reporting of an alarm event by tens of thousands of devices in a cell~\cite{METIS}.
The main concern in such scenarios is the overload of the cellular access infrastructure, i.e., the collapse of the random access channel (RACH), which happens due to the signaling overhead associated with each individual transmission~\cite{surveyRACHLTE}. We note that the RACH overload precludes any service operation, i.e., blocks the system, and it is therefore of paramount importance to prevent it.

Several methods have been recently proposed to prevent the RACH overload in LTE~\cite{3GPPR12}, in the context of M2M communications.
	Specifically, two main solutions are the extended class barring (EAB)~\cite{3GPPTS36.331} and dynamic allocation~\cite{R2104662}.
EAB is valid only for delay-tolerant M2M traffic and is an extension of the standard access class barring method. 
On the other hand, dynamic allocation is a straightforward approach: upon detection of RACH overload the number of random access opportunities (RAOs) per second is increased.
However, both schemes have inherent limitations, as they are both reactive and triggered upon RACH overload detection. Once the overload is detected, there is an additional delay until the EAB  or the dynamic allocation feedback messages are delivered from the BS to the M2M devices, which can take up to $5$~s~\cite{typical}, as these messages are typically broadcasted periodically over the paging channel.
Therefore, these two methods cannot ensure timely and reliable operation in M2M scenarios with massive synchronous arrivals, as it becomes apparent further in this text.

Motivated by the deficiencies of the reactive approaches, in this paper we propose a \emph{proactive} approach for the reliable support of M2M service. The proposed approach consists of two phases, an \emph{estimation phase} and a \emph{serving phase}, which reoccur periodically. In the first phase, the BS estimates how many M2M devices are attempting to access. We show that by using an estimator that is tuned to the LTE access mechanisms this can be done in a simple and, more importantly, fast manner, requiring just a single RAO to estimate the number of accessing users in the order of tens of thousands. Following the estimation phase, the parameters of the access mechanism are tuned such that the RAOs of the serving phase are used in an efficient way, providing a reliable service. The proposed solution can be easily incorporated in the standard LTE access mechanism, leaving the radio interfaces intact and used both for the case of massive synchronous arrivals as well as the asynchronous traffic with Poisson arrivals. In this way the mobile operators can provide M2M service in a controlled manner, with guaranteed reliability and no overload, i.e., the operators can be provided with a technical data-sheet indicating the performance of the system for a given number of devices and the associated latency. This is a significant step towards reliable M2M services in LTE, which are currently based on the best effort approach.

The rest of the paper is organized as follows. In Section~\ref{sec:lte_overview} we present a brief overview of the standard LTE RACH operation. Section~\ref{sec:lte_reenginering} is the central part of the paper, where we describe and analyze the proposed solution, as well as outline its practical implementation. Section~\ref{sec:case_study} presents a case study involving two M2M traffic classes, presenting the performance results and a comparison with competing methods for M2M access. Finally, Section~\ref{sec:conclusions} concludes the paper.

\section{LTE RACH Overview} 
\label{sec:lte_overview}

	The uplink time in LTE is divided in frames, where every frame is composed of ten subframes whose duration is 1~ms.
The LTE frequency band is organized in subcarriers, where 12 subcarriers of 15~KHz over a subframe constitute a resource block (RB). The bandwidth of LTE ranges between 6~RBs (i.e., 1~MHz) and 100~RBs (20~MHz). 
	In LTE, a random access opportunity (RAO) requires 6~RBs in a subframe.
	The number of RAOs per frame is a system parameter, ranging from one RAO every 20~subframes to one RAO every subframe.
	A typical configuration foresees one RAO every 5~ms~\cite{typical}.
	Further, up to 64 orthogonal preamble sequences are available in each RAO, which can be detected simultaneously by the base station (BS).
	The actual number of available preambles depends on the system configuration, where a typical configuration foresees 54 preambles~\cite{typical}.
	The System Information Blocks (SIB)s, where all announcements including where each RAO occurs, are broadcasted periodically via the paging procedure that occurs from every 80~ms up to every 5.12~s~\cite{3GPPTS36.331}.

	
	The LTE random access procedure, denoted as Access Reservation Procedure (ARP), consists of the following four stages.
	(1) First, a device (UE), selects one of the preambles and transmits it in one of the RAOs. 
	(2) In the case a single UE has transmitted the preamble, the eNodeB decodes it and responds by sending a random access response (RAR) message. 
	(3) This RAR message indicates the RBs where the device shall send its request consisting of a temporary ID together with the establishment cause, e.g., call, data, report, etc. 
	If two or multiple devices have selected the same preamble within the same RAO a collision occurs, the eNodeB detects this and does not send back any response.
	(4) In the last stage, the eNodeB acknowledges the information received from the device and allocates the required data resources. 
	If the UE does not receive a response to a preamble or a request, it restarts the procedure until it is successful or the maximum number of preambles retransmissions is reached.

	When the number of devices attempting access is high, most of the RACH preambles are selected by multiple devices and end in collisions.
	Consequently, most devices are not granted access and therefore retry again.	
	There reattempts coupled with the new arrivals lead to an even higher amount of attempted accesses, further overloading the RACH and with the end result of almost no device being granted access.
	The general load control mechanism in LTE is the access class barring (ACB), which works by assigning access probabilities to different access classes~\cite{3GPPTS22.011}.
	However, as the ACB does not distinguish between H2x and M2M traffic, the EAB was defined in \cite{3GPPTS36.331} to deal with potential burst of M2M traffic arrivals.
	EAB is used to explicitly restrict access from devices configured as delay tolerant.
	The core network can also trigger the admission control at the radio access network~\cite{ksentini2012cellular}, via dynamic blocking according with the load. 

	Another mechanism proposed to overcome the RACH overload is the dynamic allocation mechanism~\cite{R2104662}.
	Here, whenever the eNodeB detects the occurrence of overload, it increases the number of RAOs per frame.
	Due to the system limitations, this increase is up to one RAO per subframe, announced to the devices via the paging procedure.
	This mechanism can be further enhanced through the expansion of the LTE contention space to the code domain~\cite{ETT:ETT2656}.

\section{Proposed Solution}
\label{sec:lte_reenginering}
	\begin{figure}[tb]
		\begin{center}
			\includegraphics[width=0.8\columnwidth]{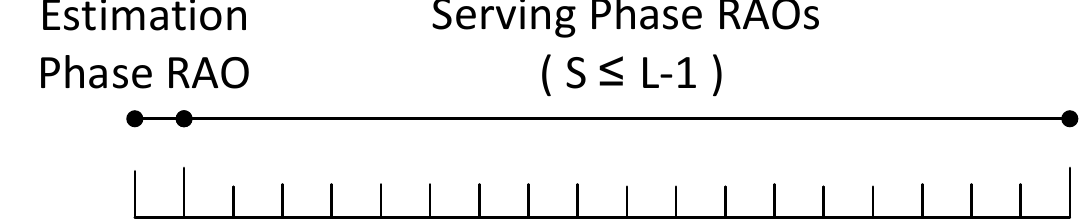}	
		\end{center}
		\caption{Proposed access frame consisting of an estimation RAO followed by $S \leq L - 1 $ serving RAOs.}
		\label{fig:frameStructure}
	\end{figure}
	The core of the proposed solution consists of a reoccurring access frame, which is composed of RAOs that are dedicated to M2M devices.\footnote{The use of dedicated resources for M2M has been proposed previously in~\cite{R2104662,6720118}, in an attempt to prevent M2M RACH accesses from affecting H2x services.}
	It is assumed that the arrival process is gated, i.e., new arrivals are accepted at the frame beginning and all arrivals during the frame wait for the beginning of the next one.
	The frame time duration is assumed to be fixed and limited to half of the maximum allowed delay $\tau$ guaranteed by the network operator.
	The frame is then composed by up to $L$ M2M dedicated RAOs within $\tau/2$\footnote{Assuming the H2x dedicated RAOs occur every 5~ms~\cite{typical}, then within a $\tau/2 = 0.5$ seconds, there will be up to $L=400$ available RAOs for M2M access, i.e., 8 RAOs per LTE frame.}. 
	Obviously, a larger $L$ implies a longer delay, but it also accommodates more devices.
	
	The frame consists of two parts, dedicated to the \emph{estimation} and \emph{serving} phase, as depicted in Fig.~\ref{fig:frameStructure}. 
	We design the estimation part such that it consists just of a single RAO and describe in Section~\ref{sub:estimationAlgorithm} the proposed estimation technique, showing that a huge range in the number of accessing M2M devices $N$ can be reliable estimated.\footnote{We note that the approach grants straightforward extension to cover the cases when the estimation phase consists of two or more RAOs.}
	The length of the serving phase $S$ is determined by the estimated number of arrivals $\hat{N}$, with the constraint that $S \leq L - 1$.
	The access algorithm in the serving $S$ is based on the standard LTE RACH operation, but tuned to $\hat{N}$ such that its resources, i.e., RAOs, are used so that the required reliability $R_{req}$ is met.
	Particularly, we distinguish two modes of operation in the serving phase.
	In the first mode, the length required by the target reliability $S_{\text{req}}$ is lower or equal to $L-1$ and the actual length is set to $S = S_{\text{req}}$.
	In the second mode, $S_{\text{req}} > L-1$, which implies that there are not enough resources to provide required service.
	In this case, the length of the serving phase is set to $S = L - 1$, and a barring factor is introduced to prevent RACH overload.
	Further details on the operation and dimensioning of the serving phase are presented in Section~\ref{sub:servingPhase}.

\subsection{Estimation Algorithm}
\label{sub:estimationAlgorithm}
		
	We assume the estimation takes place in a single RAO with $J$ preambles.
	The preambles are ordered from 1 to $J$ (in a arbitrary way) and the active devices (i.e., devices with traffic arrivals) choose one of preambles with a predefined probability.
	The probability of selecting preamble $j$ is given by:
	\begin{equation}
		p_j = \frac{p_0}{\alpha^j}, \; j=1,2,\dots, J,
	\end{equation}
	where $p_0 \leq 1$ and $\alpha>1$ are a priori determined parameters, whose choice depends on the expected range of the number of users $N$.
		
	The eNodeB observes a ternary outcome\footnote{In Section~\ref{sec:Practical_Implementation} is described how the collision detection is performed.} for each preamble - a preamble can be in the \emph{idle} state (no devices transmitted it), \emph{singleton} state (a single device transmitted it) or \emph{collision} state (two or more devices transmitted it).
	Based on the observed outcomes, the eNodeB estimates how many users are present in the frame.
	The main idea behind varying the preamble activation probability is to obtain a favorable mix of collision, singleton and idle preambles, which will allow a reliable estimation.
	The same idea is standardly used in framed slotted ALOHA-based estimation algorithms~\cite{Kodialam:2006:FRE:1161089.1161126,1433259,5370274}.
	Here we use a modification of a simple technique first proposed in~\cite{6655070}, characterized by a large estimation range.	
	The main difference with respect to~\cite{6655070} is that devices are limited to a single transmission due to the physical layer constraints.
	
\begin{figure}[tb]
		\begin{center}
			\includegraphics[width=\columnwidth]{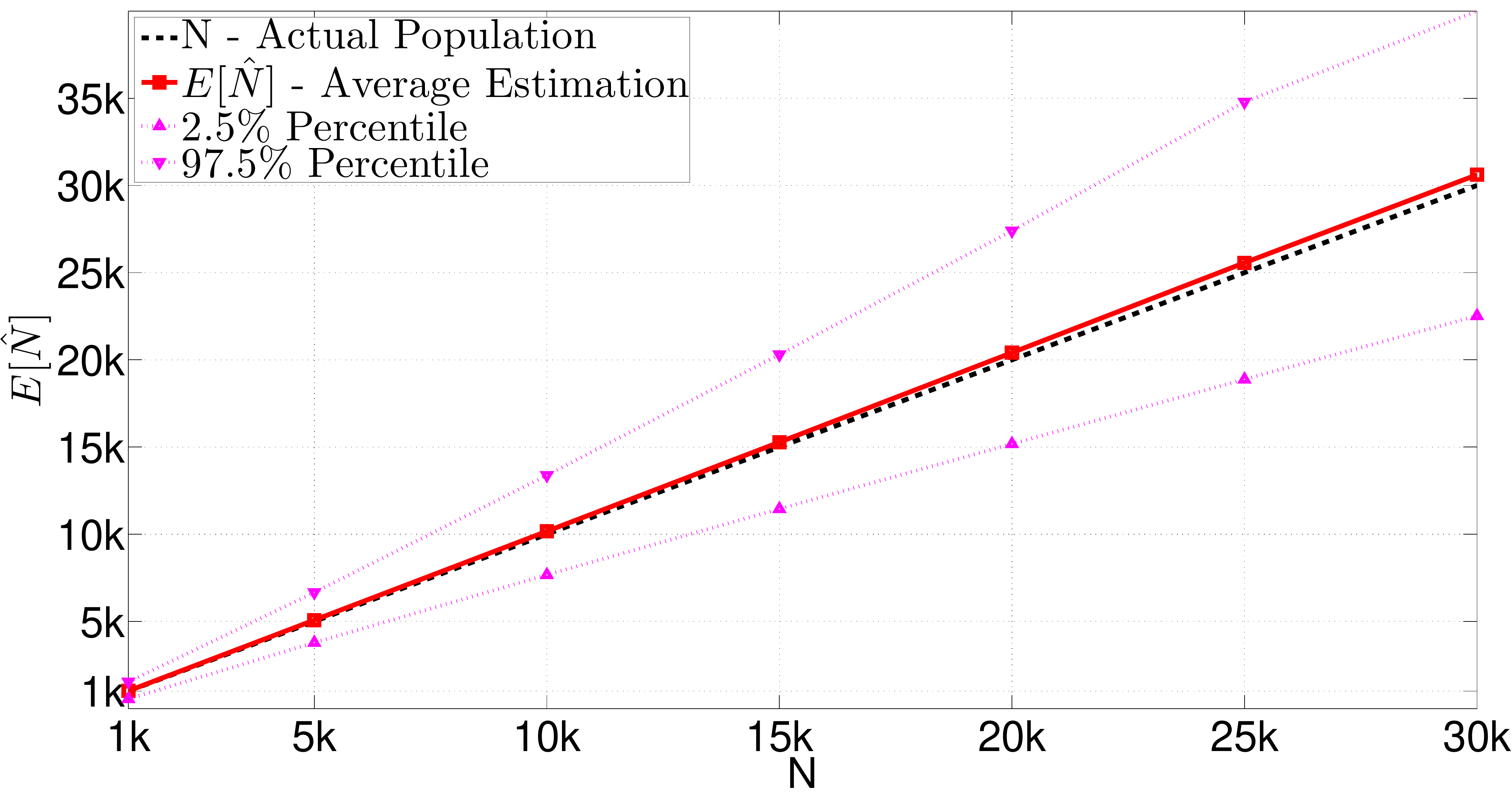}	
		\end{center}
		\caption{Proposed estimator performance when the expected arrival rate is not a priori known.
		Through exhaustive numerical search it was found that for a dynamic range between $N \in [1, 30000]$ the optimal values of the estimator parameters are $p_0 = 0.001$ and $\alpha = 1.056$.}
		\label{fig:comparisonEstimators}
	\end{figure}

	Let $a_j$ denote the probability that a device has not transmitted any of the previous $j-1$ preambles:
	\begin{equation}
		a_j = \prod_{i=1}^{j-1} (1 - p_i), \; 1 < j \leq J, 
	\end{equation}
	with initial condition $a_1 = 1$.
	Denote the observed state of the preamble $j$ preamble as $s_j$, where $s_j = 0$ if the state is idle, $s_j = 1$ if singleton and $s_j > 1$ if collision.
	The conditional probability mass function $f (s_{j} | N = n )$ is given by:
	\begin{align}
	\label{eq:approx}
		f (s_{j} | N = n ) =  \left\{
			    \begin{array}{ll} 
			      (1-p_j)^{a_j \cdot n}  &   s_{j} = 0, \\
			      a_j \cdot n \cdot (1-p_j)^{a_j \cdot n} & s_{j} = 1, \\
			      1 - \left[ 1  +  a_j \cdot n\right] \cdot (1-p_j)^{a_j \cdot n} & s_{j} > 1.\\
			    \end{array}
			  \right.
	\end{align}
	We note that the above expression is an approximation, as it assumes only the expected number of users capable of transmitting preamble $j$, i.e., $a_j \cdot N$.
	However, this approximation allows for an elegant solution that yields accurate results, as demonstrated further.
	The estimation of $N$ is performed using the sequence of observations $\{s_j, j = 1,\dots,J\}$, using the maximum likelihood approach:
		\begin{align}
		\label{eq:prod}
		  \hat N & = \arg \max_{n} \prod_{j=1}^{J} f (s_{j} | N = n ) \nonumber \\
		  		 & = \arg \max_{n} \sum_{j=1}^J \ln f (s_{j} | N = n ),
		\end{align}
		which is obtained by solving for $n$ the following equation:
		\begin{align}
			 \frac{\partial }{\partial n} \left( \sum_{j} \ln f (s_{j} | N = n ) \right) &= 0, 
			\label{eq:maximize}
		\end{align}
	using a root-finding method.
	We conclude by presenting the estimator performance in Fig.~\ref{fig:comparisonEstimators}, where it can be observed that the $E[\hat N] $ follows closely the actual value of $N$.
	%
	%
	

	\subsection{Serving Phase}
	\label{sub:servingPhase}
		
\begin{figure}[tb]
			\begin{center}
				\includegraphics[width=0.75\columnwidth]{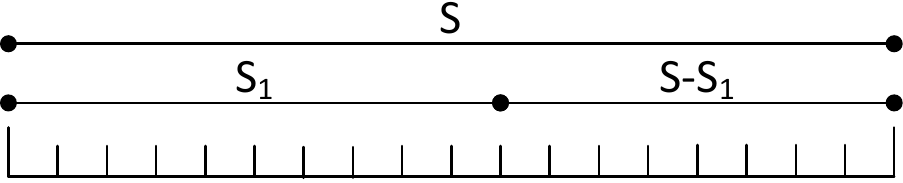}
			\end{center}
			\caption{Serving scheme structure, where the serving phase is composed by $S$ RAOs grouped into two frames of lenght $S_1$ and $S-S_1$.}
			\label{fig:OneShotAndTwoPoolsAccess}
		\end{figure}

		The number of RAOs in the serving phase $S$ should be, if possible, dimensioned according to $N$ such that the required reliability $R_{req}$ is met.
		On the other hand, $S$ also depends on the access scheme employed in the serving phase, which is based on the LTE RACH operation, i.e., based on framed slotted ALOHA.
		In the further text, we assume that the serving phase consists of two frames, as depicted in Fig.~\ref{fig:OneShotAndTwoPoolsAccess}.
		In the first frame the devices attempt access by transmitting a single randomly selected preamble in a randomly selected RAO, while in the second frame all devices that collided in the first frame reattempt access in the same way.\footnote{We assume that the number of preambles is constant for all RAOs and equal to $J$.}
		We show that in this way we can achieve close-to-one reliability for a huge range of accessing devices.\footnote{In principle, it could be argued that variants in which more than one retransmission per collided device is allowed could provide a higher reliability with the same number of RAOs. However, we demonstrate that the proposed approach shows rather favorable performance and allows for tractable modeling and analysis.}

		We define reliability $R(N)$ as the probability of a device successfully obtaining a data resource when there are $N$ contending device, at the completion of the Access Reservation Procedure discussed in Section~\ref{sec:lte_overview}.
		For this to occur, the device has to be the only one to select a preamble from the RAOs available in either of the frames in the serving phase. We then model the reliability $R(N)$ as:
		\begin{equation}
		\label{eq:ReliabilityScheme2}
			R(N) = P_1(N) + \left[ 1 - P_1(N)\right]  P_2(N),
		\end{equation}
		where $P_1(N)$ and $P_2(N)$ denote the probabilities that a device does not collide in the first and second frames, respectively.
		In the first frame, the success probability is the probability that a device is the only one to select one of $J$ preambles in one of $S_1$ RAOS, when there are $N$ contending devices, which is:
		\begin{equation}
			P_{1}(N) = \left(1 - \frac{1}{S_1  J}\right)^{N-1}.
		\end{equation}
		The success probability in the second frame depends on the number of collisions $N_C$ in the first frame.
		Denote by $\text{Pr}[ N_C =k | N, S_1 ]$ the probability mass function (pmf) of the number of collisions in the first frame, conditioned on $N$ and $S_1$, where:
		\begin{equation}
		\label{eq:c}
			\text{Pr} [ N_C=k | N, S_1] =  \text{Pr} [ N_S = N-k | N, S_1],
		\end{equation}
		where $N_S$ denotes the number of successful devices in the first frame.
		The pmf $\text{Pr} [ N_S = N-k | N, S_1 ]$ can be modeled as a balls and bins problem, where the balls and bins represent respectively the devices and the contention resources (i.e., preambles and RAOs).
		In~\cite{vogt2002efficient} this distribution is provided in a closed form expression as follows:
		\begin{equation}
			\text{Pr} [ N_S = s | N, S_1] = \frac{\binom{S_1 J}{s} \prod_{k=0}^{s-1} (N-k) G(S_1 J-s,N-s)} {(S_1 J)^{N}},
		\end{equation}
		where:
		\begin{equation}
			G(u,v) = u^v + \sum_{t=1}^{v} (-1)^t \prod_j^{t-1}[(v-j)(u-j)](u-t)^{v-t}\frac{1}{t!}.
		\end{equation}
		The probability of a device being successful in the second frame $P_{2}(N)$, from the law of total probability, is given by:
		\begin{equation}
		\label{eq:eq}
			P_{2}(N) = \sum_{k=2}^{N} \left(1-\frac{1}{(S-S_1)J}\right)^{k-1} \cdot \text{Pr} [ C=k | N, S_1 ].
		\end{equation}
		Using \eqref{eq:ReliabilityScheme2}, \eqref{eq:c} and \eqref{eq:eq} it is possible to find the optimum $S_1$ that maximizes~\eqref{eq:ReliabilityScheme2} and the minimum $S_{req}$ that meets $R_{req}$ through a numerical search.
		
		When the number of required contention resources is higher than the maximum available $S_{\text{req}} > L-1$, a barring factor $Q$ is introduced. This barring probability is then used independently by each device in a Bernoulli trial with probability $1-Q$ to decide if the device should attempt to access the serving phase.
		To account with the barring probability,~\eqref{eq:ReliabilityScheme2} is redefined as follows:
		\begin{align}\label{eq:ServingSpaceScheme2WithBarring}
			R_Q(N,Q) = (1-Q) \sum_{k=0}^{N-1} B(N-1,k,Q) R(k)
		\end{align}
		where $B(x,y,z) =\binom{x}{y} (1 - z)^y z^{x - y}$ is the binomial pmf. The optimal $Q$ that maximizes~\eqref{eq:ServingSpaceScheme2WithBarring} is found via:
		\begin{equation}\label{eq:MaxizimizationScheme2Q}
			\operatorname*{arg\,max}_{Q}  R_Q(N,Q) := \left\{Q \mid \forall y : R_Q(N,y) \leq R_Q(N,Q)\right\}.
		\end{equation}

		We summarize the dimensioning of the serving phase in Algorithm~\ref{alg:contentionResourceAllocation}.
		\begin{algorithm}[h]\label{alg:contentionResourceAllocation}
	 		Input {$\hat{N}$, $R_{req}$, $L$}; \\
	 		$S_{req}$ is computed from \eqref{eq:ReliabilityScheme2}\;
	 		\eIf{$S_{req} \leq (L-1)$}{
	   			$Q = 0$; $S = S_{req}$\;
	   		}{
	   			$Q$ is computed from \eqref{eq:MaxizimizationScheme2Q}; $S = L-1$\;
	  		}
			Output {$S$, $Q$}; \\
	 		\caption{Dimensioning of the size of the serving phase frame $S$ and the associated barring probability $Q$.}
		\end{algorithm}

\subsection{Practical Implementation}\label{sec:Practical_Implementation}

	All the information required by the devices to attempt access is broadcasted, similarly to the EAB, in a new system information message (SIB)~\cite{TR37.8682011} that takes place in each access frame, immediately after the estimation RAO.
 	This SIB message includes the following information:
	First it indicates in which subframe the upcoming estimation RAO will take place together with the values of $p_0$ and $\alpha$ and the number of preambles $J$.
	Further, it informs the contending devices of the number of RAOs in the serving phase and $S_1$. Finally, a bitmap is included which indicates in which subframes these RAOs will occur.
	If the load exceeds the amount of capacity pre-reserved by the operator, the barring factor $Q$ is also included in the SIB, to prevent the RACH overload.

	The proposed scheme operation is then as follows.
	Assume that $N$ contending devices become active prior to start of the access frame.
	When the estimation RAO occurs, each of these $N$ devices attempt access, according with the procedure defined in Section~\ref{sub:estimationAlgorithm}, enabling the eNodeB to obtain the estimation of the number of arrivals $\hat{N}$.
	The detection of collisions in the estimation phase, is performed during the execution of the Access Reservation Procedure. Namely, after the devices that have selected the same random access preamble, transmit their UE request, which will result in a collision as described in Section~\ref{sec:lte_overview}. 
	Based on $\hat{N}$, the eNodeB then defines how many RAOs are required in the serving phase to reach the contracted $R_{req}$ and informs the devices where these RAOs will occur by broadcasting the corresponding SIB.
	Then, the contending devices select randomly between the serving RAOs, using the ARP described in Section~\ref{sec:lte_overview}.
	In the meantime, other contending devices become active, which will wait until the start of the next access frame before proceeding in the same way.

	We note that the proposed scheme requires minimal changes to the current LTE protocol, with no modifications to the physical layer at all.

\section{Case Study for Two M2M Traffic Classes}
\label{sec:case_study}
	\begin{figure}[tb]
		\begin{center}
			\includegraphics[width=\columnwidth]{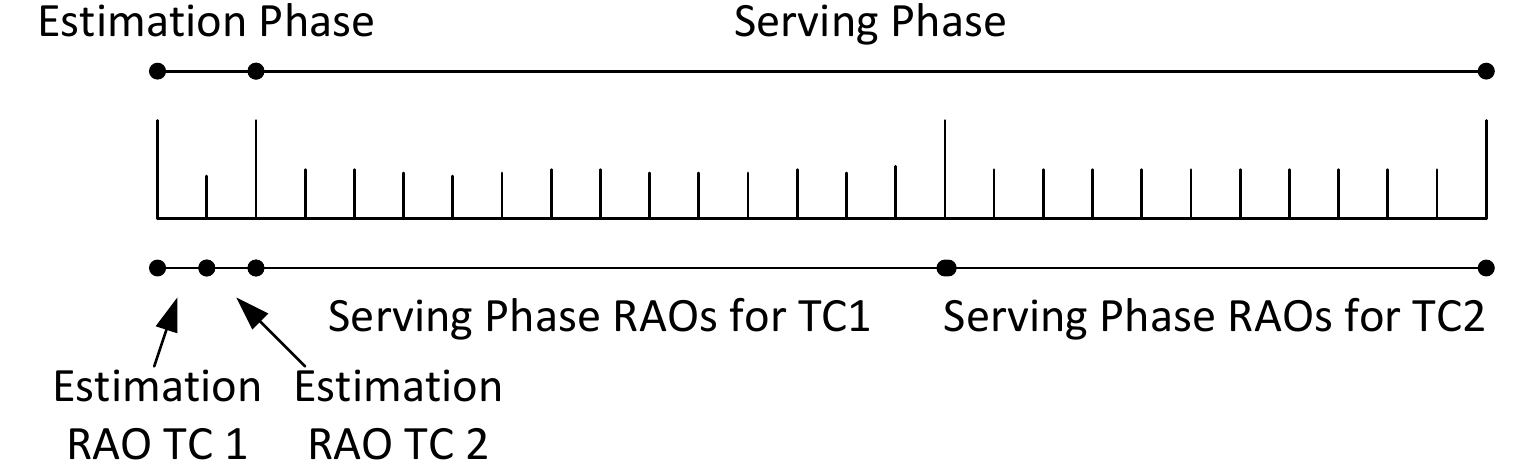}
		\end{center}
		\caption{Proposed access scheme for two traffic classes TC1 and TC2, where TC1 has priority over TC2.}
		\label{fig:TwoTrafficClasses}
	\end{figure}

	We now consider a case study with two traffic classes characterized by different requirements and serving probabilities.
	Let traffic class 1 (TC1) and traffic class 2 (TC2), have a respective reliability requirement $R^{(1)}_{req}$ and $R^{(2)}_{req}$. 
	Further, let TC1 have priority access to the available serving RAOs over TC2, e.g., alarm reports take priority over periodic reporting in the context of smart metering.
	Specifically, we try first to reach as close as possible to $R^{(1)}_{req}$ and only then as close as possible to $R^{(2)}_{req}$.
	Furthermore, we assume that each class has separate estimation and serving phases, as depicted in Fig.~\ref{fig:TwoTrafficClasses}.
	At the beginning of the frame there is one estimation RAO for each traffic class, where the number of contending devices of each class is estimated to be $\hat{N}_1$ and $\hat{N}_2$ respectively.
	With the knowledge of $\hat{N}_1, \hat{N}_2$ we define a resource allocation strategy based on the scheme described in Section~\ref{sec:lte_reenginering}.
	
	The access frame duration -- demarcated by the estimation phase RAOs occurrence -- is constrained by the traffic class with the most stringent latency requirement, here given by TC1.
	Although, in this study we consider that both TCs have an estimation phase in each access frame, we note that in the case where TC2's latency requirement is much larger than TC1's, it might be worthwhile to consider the case where TC2 estimation RAO only occurs in some of the access frames, in order to optimize the amount of RAOs dedicated for estimation.

	\subsection{Serving Phase Size and Barring Factor for Two Traffic Classes}
	\label{sub:servingPhaseTwoClasses}

		The extension of the analysis in Section~\ref{sec:lte_reenginering} to two traffic classes is straightforward.
		Denoting as $S^{(1)}$ and $S^{(2)}$ the amount of serving RAOs respectively required to serve TC1 and TC2 to meet the reliability requirements of each class, $R^{(1)}_{req}$ and $R^{(2)}_{req}$. 
		The main distinction from the case with a single traffic class, is that now there are three different operation regimes: (i) $S^{(1)}_{req} + S^{(2)}_{req} \leq L-2$; (ii) $S^{(1)}_{req} + S^{(2)}_{req} > L-2$ with $S^{(1)}_{req} < L-2$; and (iii) $S^{(1)}_{req} > L-2$.
		In (i) each traffic class receives the number of required serving RAOs.
		In (ii) a barring factor $Q^{(2)}$ is introduced to the lower priority class TC2, while no barring is necessary for the high reliability class TC1.	
		Finally in (iii), TC2 is completely barred ($Q^{(2)}=1$) and a barring factor $Q^{(1)}$ is introduced for the high reliability class TC1.
		This procedure is described in detail in Algorithm~\ref{alg:contentionResourceAllocationTwoClasses}.
		\begin{algorithm}[h]
		\label{alg:contentionResourceAllocationTwoClasses}
 			Input {$\hat{N}_1$, $\hat{N}_2$, $R^{(1)}_{req}$, $R^{(2)}_{req}$, $L$}; \\
 			$S^{(1)}_{req}$ and $S^{(2)}_{req}$  computed from \eqref{eq:ReliabilityScheme2}\;
 			\uIf{$S^{(1)}_{req} + S^{(2)}_{req} \leq L-2$ }{
				$Q^{(1)} = Q^{(2)} = 0$; $S^{(1)} = S^{(1)}_{req}$; $S^{(2)} = S^{(2)}_{req}$\;
			}
			\uElseIf{$S^{(1)}_{req} + S^{(2)}_{req} > L-2$ and $S^{(1)}_{req} < L-2$}{
				$Q^{(1)} = 0$; $S^{(1)} = S^{(1)}_{req}$\;
				$S^{(2)} = L - 2 - S^{(1)}$; $Q^{(2)}$ computed from \eqref{eq:MaxizimizationScheme2Q} \;
   			}
			\Else{
				$Q^{(1)}$ computed from \eqref{eq:MaxizimizationScheme2Q}; $S^{(1)} = L-2$\;
				$S^{(2)} = 0$; $Q^{(2)} = 1$\;
			}
			Output {$S^{(1)}$, $S^{(2)}$, $Q^{(1)}$, $Q^{(2)}$}; \\ 			
 			\caption{Dimensioning of the size of the serving phase frame $S^{(1)}$ and $S^{(2)}$ and associated barring probabilities $Q^{(1)}$ and $Q^{(2)}$.}
		\end{algorithm}

		\subsection{Performance Results and Discussion} 
		\label{sec:performance_results}
		\begin{table}
			\centering
			\caption{Legacy LTE system parameters.}
			\begin{tabular}{@{}ll|ll@{}}
				\toprule
				\textbf{Parameter}       & \textbf{Value}  	& \textbf{Parameter}  & \textbf{Value}  \\ \midrule
				Preambles per RAO ($J$)  & 54  				& MSG~2 Window  	  & 5~ms  \\ 
				Max. RAOs per LTE frame  & 8  		    	& MSG~4 Timer   	  & 24~ms \\ 
				Max. Retransmissions     & 9  			    & Contention Timer    & 48~ms  \\ 
				System BW			     & 20~MHz		    & Backoff 			  & 20~ms \\ 
				eNodeB Processing Time   & 3~ms 			& UE Processing Time  & 3~ms  \\ 
				\bottomrule
			\end{tabular}
		\label{table:parameters}
		\end{table}
		The performance results are obtained from a LTE event-driven simulator implemented in MATLAB, which models the complete access reservation procedure described in Section~\ref{sec:lte_overview}.
		For the same network conditions, we compare the performance of the legacy LTE with dynamic allocation\footnote{We do not include a numerical comparison with EAB, as the algorithm that controls the blocking of M2M traffic is not standardized.} with the performance of the proposed scheme.
		The system parameters of interest for the legacy system are listed in Table~\ref{table:parameters}; we assume an ideal, best-case dynamic allocation, where the network overload is detected instantaneously and there is no delay to change the parameters of the system such as the number of available RAOs.
		The incoming traffic is classified into two traffic classes: (TC1) alarm and (TC2) periodic reporting; where the alarm reporting takes priority over periodic reporting.

		The alarm reporting case is modeled by a Beta distribution with parameters $\alpha = 3$ and $\beta = 4$ \cite{Madueno2017}, which trigger $N_1$ smart meters within the cell to access the same access frame with latency requirement $\tau_1$.
		The periodic reporting is modeled as a Poisson process with total arrival rate $\lambda = N_2/RI$, where $N_2$ denotes the number of M2M devices and $RI = \tau_2 = 60$ s, chosen so to match the arrival rate and latency requirement $\tau_2$ of a typical M2M application such as smart metering~\cite{Madueno2017}.
		
		The performance comparisons are done using different access frame $L$ lengths, obtained from half of the maximum allowed delay for alarm reporting $\tau_1/2 = \{ 0.5, 2.5, 5 \}$ seconds.\footnote{Thus, taking into account the 2~RAOs per frame reserved for other purposes (e.g., H2x), the maximum amount of RAOs in each frame is then $L=\{ 400, 2000, 4000\}$.}
		\begin{figure}[tb]
			\begin{center}
				\includegraphics[width=\columnwidth]{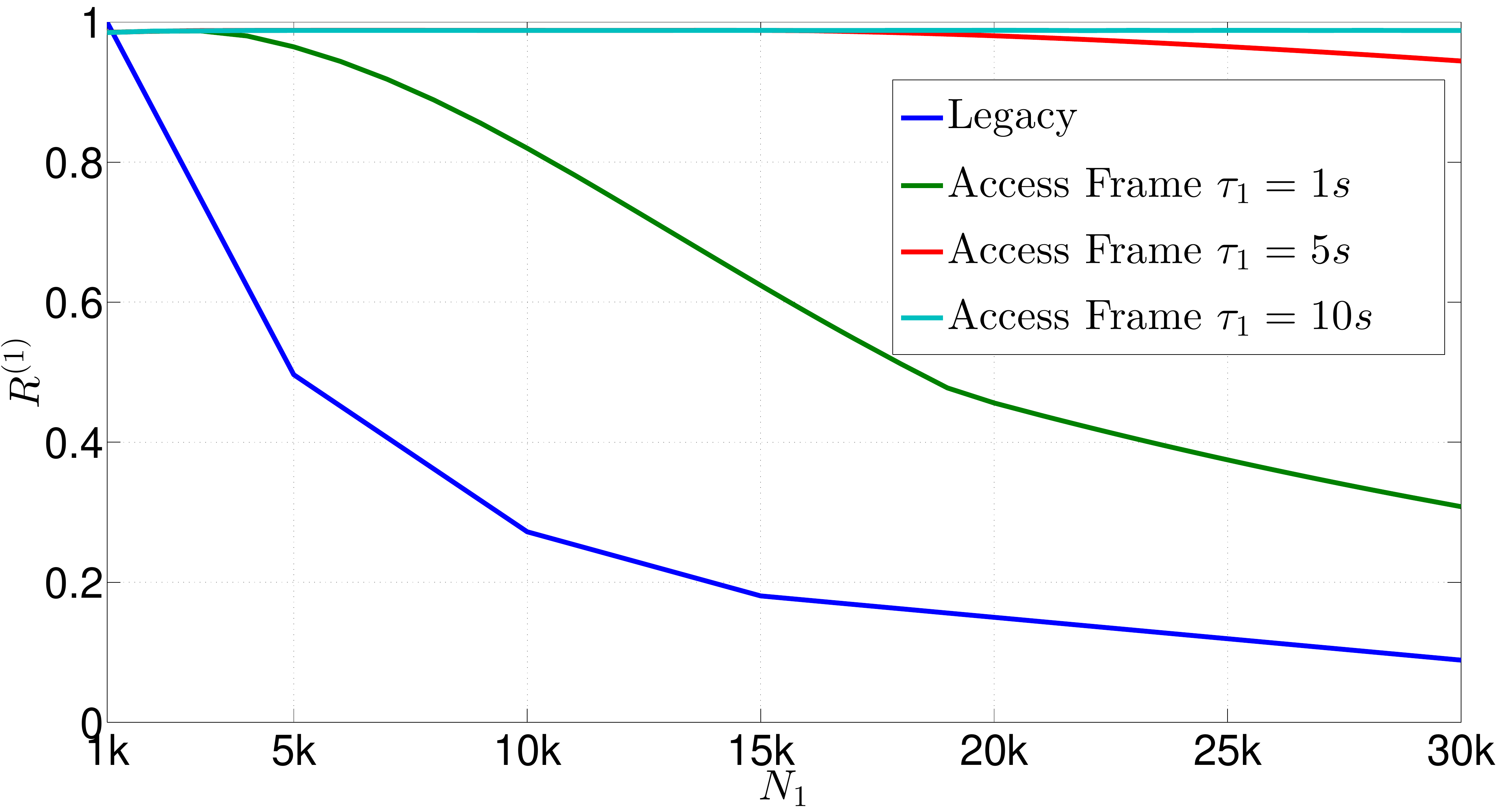}
			\end{center}
			\caption{Achievable transient $R^{(1)}$ within the access frame by the legacy and access frame solutions, with $N_2 = 10k$ and $R^{(1)}_{req} = R^{(2)}_{req} = 0.99$.}
			\label{fig:performanceTC1}
		\end{figure}
		The performance evaluation is performed with the focus on the reliability achieved within the duration of the access frame.
		Specifically, we illustrate the performance during the peak of traffic due to the alarm reporting.
		The achievable reliability of TC1, $R^{(1)}$, for different number of active TC1 devices is shown in Fig.~\ref{fig:performanceTC1}.
		We first observe that the LTE legacy with dynamic allocation, is not able to provide reliable access for $N_1 > 1k$ (in the legacy solution TC1 and TC2 are treated in the same way).
		On the other hand, the proposed mechanism is able to provide a reliable service for a considerably higher range of simultaneously accessing devices.
		Specifically, the proposed scheme provides service with a reliability guarantee of $R^{(1)}_{req} = 0.99$ for up to $N_1 = 30k$ smart meters if the tolerable delay is $\tau_1 = 10$~s.	
		For TC2, the offered reliability will be constrained by the amount of TC1 arrivals in the same access frame. However, due to TC1 bursty nature and the less restrictive TC2 latency requirement (i.e. $\tau_1 < \tau_2$), we have observed that, after the ``storm'' caused by the alarms is over, our solution is able to met the set $R^{(2)}_{req}$.

		We emphasize, that beyond this specific example, our proposed solution is tailored to offer the traffic reliability requirements, as long as the allowed latency constraints are in accordance with the number of devices to be served. Furthermore, it enables to achieve a trade-off between latency and reliability.

\section{Conclusions}
\label{sec:conclusions}

	One of the key challenges associated with machine-to-machine (M2M) communications in cellular networks is to be able to offer service with reliability guarantees, particularly when a massive amount of simultaneous M2M arrivals occurs.
 	While current solutions take a reactive stance when dealing with massive arrivals, by either imposing barring probabilities or increasing the contention space, they do so without knowledge of the volume of incoming traffic.

	Here we propose a proactive approach, based on dedicated access resources for the M2M traffic, combined with a novel frame based serving scheme composed by an estimation and a serving phase.
	In the estimation phase the volume of arrivals is estimated and then used to dimension the amount of resources in the serving phase, such that reliable service guarantees are provided.
	The provided framework can be extended for more than two traffic classes, which is one of the future work directions.
	Other directions include combination of the proposed approach with the existing access control mechanisms, such as the EAB.

\section*{Acknowledgment}
The research presented in this paper was partly supported by the Danish Council for Independent Research (Det Frie Forskningsr\aa d), grants no. 11-105159 ``Dependable Wireless Bits for Machine-to-Machine (M2M) Communications'' and no. DFF-4005-00281 ``Evolving wireless cellular systems for smart grid communications'', and partly funded by the EU project SUNSEED, grant no. 619437.

\bibliographystyle{IEEEtran}

\end{document}